# Wafer-scale synthesis and transfer of graphene films

Youngbin Lee[1], Sukang Bae[1], Houk Jang[2], Sukjae Jang[1], Shou-En Zhu[2], Sung Hyun Sim[3], Young Il Song[4], Byung Hee Hong[1,3*] & Jong-Hyun Ahn[1,2*]

*[1]SKKU Advanced Institute of Nanotechnology (SAINT) and Center for Human Interface Nano Technology (HINT), [2]School of Advanced Materials Science and Engineering, [3]Department of Chemistry, Sungkyunkwan University, Suwon 440-746, Korea. [4]Products Development Group, Digital Solution R&D Team, Digital & IT Solution Division, Samsung Techwin, Seongnam 462-807, Korea*

[*]Corresponding Authors: *ahnj@skku.edu* and *byunghee@skku.edu*

*Abstract*

We developed means to produce wafer scale, high-quality graphene films as large as 3 inch wafer size on Ni and Cu films under ambient-pressure and transfer them onto arbitrary substrates through instantaneous etching of metal layers. We also demonstrated the applications of the large-area graphene films for the batch fabrication of field-effect transistor (FET) arrays and stretchable strain gauges showing extraordinary performances. Transistors showed the hole and electron mobilities of the device of $1,100 \pm 70$ cm$^2$/Vs and $550 \pm 50$ cm$^2$/Vs at drain bias of -0.75V, respectively. The piezo-resistance gauge factor of strain sensor was ~6.1. These methods represent a significant step toward the realization of graphene devices in wafer scale as well as application in optoelectronics, flexible and stretchable electronics.

Graphene and related materials have been intensively studied due to their fascinating electrical and mechanical properties.[1-4] The recent advances in large-scale synthesis of graphene films by chemical vapour deposition (CVD) on Ni layers are expected to enable various macroscopic applications such as transparent conducting films useful for flexible/stretchable electronics.[5-7] However, the lack of efficient etching and transfer methods practically limited the scale of production, since, for instance, the etching time for Ni layers increases exponentially with the size of graphene films. On the other hand, the low-pressure growth of graphene on Cu foils is known to be advantageous in terms of controlled thickness and quality, but the included vacuum process would be unfavourable for cost and time-effective production.[8] Although the CVD method is advantageous for large-area growth of graphene, inevitably it requires a rigid substrate that can stand the high temperature above ~900 °C as well as an etching process for removing catalyst layers, which precludes the direct use of graphene on as-grown substrates or the use of polymer substrates at low temperatures.[8] Therefore, the transfer of graphene films onto a foreign substrate is an essential process particularly for flexible/stretchable electronics based on polymers.[9-15] However, the transferrable size of graphene has been limited below a few centimeter scale due to the size limit of rigid substrates as well as the inhomogeneity of reaction temperature inside a CVD furnace.

Here we present a wafer scale ambient-pressure growth of high-quality graphene films as large as 3 inch wafer size on Ni and Cu films, followed by instantaneous etching of metal layers and polymer-supported transfer onto arbitrary substrates. This large area synthesis and transfer methods provide improved scalability and processibility of graphene films ready for use in wafer scale devices and flexible/stretchable electronics. We also demonstrate the applications of the large-area graphene films for the batch fabrication of field-effect transistor (FET) arrays and stretchable strain gauges showing extraordinary performances.

Figure 1 shows a schematic illustration of the fabrication steps. First, the 3 inch $SiO_2$/Si substrates, coated with 300nm-thick Ni or 700nm-thick Cu are inserted to a tubular quartz tube and then heated up to 1000°C under ambient pressure with flowing $H_2$ and Ar (or He). After flowing reaction gas mixtures ($CH_4$ : $H_2$ : Ar = 250 : 325 : 1,000 sccm for Ni and $CH_4$ : $H_2$ : He = 50 : 15 : 1,000 sccm for Cu) for ~5 min, the sample is rapidly cooled down to room temperature. The average number of graphene layers grown on a Ni catalyst ranged from 3 to 8, depending on the reaction time and cooling rates. On the other hand, the mono and bi-layer graphene grows predominantly on a Cu catalyst.

As a method of producing graphene devices in wafer scale, we develop a transfer method that can instantly etch metal layers. The polymer supports such as soft polydimethylsiloxane (PDMS) stamps[12] and thermal-release tapes (Nitto Denko Co.)[16] are attached to the graphene films grown on metal layers. The supports adhered to the substrate are then soaked in water. After a few minutes, the support/graphene/metal layers are detached from $SiO_2$ by water intervening between metal and $SiO_2$. A gentle ultrasonication enhances the penetration rate of water. The separated support/graphene/metal layers are soaked with $FeCl_3$ solution to remove metal layers, and then the resulting graphene film on the polymer support is ready to be transferred onto arbitrary substrates. This approach needs just a few minutes to remove metal layers completely, while the previous wafer-size etching may take a few days.[6,7] Next, the transfer printing delivers these films onto a polyethylene terephthalate (PET) film or a rubber substrate. An additional adhesive layer such as photo-curable epoxy films is helpful for improving the transfer performance. Finally, conventional photolithography and reactive ion etching (RIE) with $O_2$ plasma are employed to pattern the graphene films for device applications.[17-19] Alternatively, the graphene on metal/$SiO_2$/Si wafers can be pre-patterned in the same way before detaching and etching of metal layers.

Figure 3a show some of representative photographs of graphene films grown on a 3 inch wafer. The wafer scale graphene films can be transferred on arbitrary substrates without altering their properties or layout. Figure 3b displays an optical image of these films printed on a transparent plastic sheet, which was positioned above a logo of SKKU and Samsung to illustrate the level of optical transparency. Usually, the wafer-scale graphene growth on Ni layers hardly produces monolayer graphene, but it is advantageous in terms of patterned growth for microelectronics.[7] The outstanding mechanical and optical properties may create a possibility of using the large-area graphene as flexible transparent electrodes[8]. In order to demonstrate such capability, we also transfer the graphene films on flexible PET and stretchable elastomeric PDMS substrates (Figure 2c,d), showing excellent electromechanical modulation as we will discuss later. By using the pre-patterned graphene on the metal layer, we transferred various sizes and shapes of graphene film to a flexible substrate. The three-element rosette strain gauge is an example of large-area pattern (Fig. 2e).

Figure 3 shows that the Raman spectra of the graphene films grown on Cu and Ni layers and transferred on PDMS substrates. Here, red and blue colors denote those of thin graphene films grown on Cu film, obtained from the corresponding colored spots in the inset optical microscope image. These indicate that the graphene films synthesized on Cu films are dominantly mono- or bi-layers with small D-band peaks indicating the high-quality of graphene structures. On the other hand, the spectrum of graphene films grown on Ni indicates the property of multilayer (black color). After transfer these films onto PDMS, the intensity of D band peak in spectrum was not significantly changed, which indicates that the films can be transferred on other substrates without alerting their quality.[20]

The sheet resistance of the graphene film with 95% transparency is ~510 Ohm/square, while the film with 80% transmittance exhibits ~280 Ohm/square (Figure 3b). The large-scale graphene synthesis and transfer-printing methods enable the fabrication of wafer-scale device arrays through conventional photolithography processes. Thus, we fabricate the arrays of FETs on a 3-inch $SiO_2$/Si wafer using the patterned graphene films as channel materials. Figure 4a shows an image of device arrays (~16,200 devices) on $SiO_2$/Si wafer. The inset indicates the optical image of a patterned graphene film located in channel. Figure 4b presents the transfer characteristic of a representative device with channel length ($L_C$: 5μm) and channel width ($L_W$: 5μm) and inset provides a schematic cross section of a FET. The hole and electron mobilities of the device are estimated to be 1,100 ± 70 $cm^2$/Vs and 550 ± 50 $cm^2$/Vs at drain bias of -0.75V, respectively. In addition, we show the application of printed graphene films for strain sensors (Figure 4c,d). The zigzag type graphene electrodes of 300 μm wide and 140 mm long conducting paths are patterned on a PDMS substrate (Figure 4c). The resistance increases from ~492kΩ to ~522kΩ with applied strain up to 1%. The resistance change is reproducible even after hundreds of repetitions. The gauge factor (*GF*) of the strain sensor can be calculated using following equation

$$GF = \frac{\Delta R}{R} / \varepsilon$$

where *ΔR/R* and *ε* denotes a relative resistance change and the strain induced to gauge, respectively. The corresponding piezo-resistance gauge factor is 6.1, which is much better than that of conventional strain gauges based on metal alloys.[21]

In summary, we have demonstrated a promising route to synthesize and transfer wafer scale graphene films that are highly conducting and transparent. Developing scalable, high-throughput transfer methods of graphene films from as-grown rigid substrate to more useful, large area flexible/stretchable substrates would realize the practical use of graphene transparent electrodes for optoelectronic applications such as solar cells, touch sensors and related flexible electronics in the future.

**Acknowledgements** We thank K. S. Kim for assisting in electrical measurements, H. K. Kim for assisting in graphene synthesis. This work was supported by the National Research Foundation of Korea(NRF) funded by the Ministry of Education, Science and Technology (2009-0082608, Basic Science Research Program 2009-0083540)  and Research Center of Break-through Technology Program through the Korea Institute of Energy Technology Evaluation and Planning (KETEP) funded by the Ministry of Knowledge Economy (2009-3021010030-11-1).

# References

(1) Geim, A. K.; Novoselov, K. S. *Nature Mater*. **2007**, *6*, 183–191.

(2) Novoselov, K. S.; Geim, A. K.; Morozov, S. V.; Jiang, D.; Katsnelson, M. I.; Grigorieva, I. V.; Dubonos, S. V.; Firsov, A. A. *Nature* **2005**, *438*, 197–200.

(3) Zhang, Y.; Tan, J. W.; Stormer, H. L.; Kim, P. *Nature* **2005**, *438*, 201–204.

(4) Wang, X.; Zhi, L.; Müllen, K. *Nano Lett* **2008,** *8,* 323–327.

(5) Yu, Q.; Lian, J.; Siriponglert, S; Hao, L; Yong P. C.; Shin-Shem P. *Appl. Phys. Lett.* **2008**, *93*, 113103.

(6) Reina, A.; Jia, X.; Ho, J.; Nezich, D.; Son, H.; Bulovic, V.; Dresselhaus, M. S.; Kong, J. *Nano Lett*. **2008**, *9*, 30-35.

(7) Kim, K. S.; Zhao, Y.; Jang, H.; Lee, S. Y.; Kim, J. M.; Kim, K.S.; Ahn, J. H.; Kim, P.; Choi, J. Y.; Hong, B. H. *Nature* **2009**, *457*, 706-710.

(8) Li, X.; Cai, W.; An, J.; Kim, S.; Nah, J.; Yang, D.; Piner, R.; Velamakanni, A.; Jung, I.; Tutuc, E.; Banerjee, S. K.; Colombo, L.; Ruoff, R. S. *Science* **2009**, *324*, 1312-1314.

(9) Zhou, Y.; Hu, L.; Grüner, G. *Appl. Phys. Lett*. **2006**, *88*, 123109.

(10) Ishikawa, F. N.; Chang, H. K.; Ryu, K.; Chen, P. C.; Badmaev, A.; Arco L. G. D.; Shen, G.; Zhou, C.; *ACS Nano* **2009**, *3*, 77-79.

(11) Allen, M. J.; Tung, V. C.; Gomez, L.; Xu, Z.; Chen, L. M.; Nelson, K. S.; Zhou, C.; Kaner, R. B.; Yang, Y. *Adv. Mater*. **2009**, *21*, 2098-2102.

(12) Reina, A.; Son, H.; Jiao, L.; Fan, B.; Dresselhaus, M. S.; Liu, L. F.; Kong J. *J. Phys. Chem. C* **2008**, *112*, 17741-17744.

(13) Chen, J.-H.; Ishigami, M.; Jang, C.; Hines, D. R.; Fuhrer, M. S.; Williams, E. D. *Adv. Mater*. **2007**, *19*, 3623-3627.

(14) Kang, S.J.; Kocabas, C.; Kim, H.S.; Cao, Q.; Meitl, M. A.; Khang, D.Y.; Rogers, J. A. *Nano Lett*. **2007**, *7*, 3343-3348.

(15) Liang, X.; Fu, Z.; Chou, S. Y. *Nano Lett*. **2007**, *7*, 3840-3844.

(16) Song, L.; Ci, L.; Gao, W.; Ajayan, P. M. *ACS Nano* **2009**, *3*, 1353-1356.

(17) Meric, I.; Han, M. Y.; Young, A. F.; Ozyilmaz, B.; Kim, P.; Shepard, K. L. *Nature Nanotech*. **2008**, *3*. 654-659.

(18) Echtermeyer, T. J.; Lemmea, M. C.; Bausa, M.; Szafraneka, B.N.; Geimb, A. K.; Kurza, H. *IEEE Electro. Dev. Lett*. **2008**, *29*, 952-954.

(19) Lin, Y.-M.; Jenkins, K.A.; Valdes-Garcia A.; Small, J. P.; Farmer, D. B.; Avouris, P. *Nano Lett*. **2009**, *9*, 422-426.

(20) Wang Y. Y, et al *J. Phys. Chem. C* **2008,** *112,* 10637–10640.

(21) Chriac, H.; Urse, M.; Rusu, F.; Hison, C.; Neagu, M. *Sensor and Actuators A* **1999**, *76*, 376-380.

**Figure captions**

**Figure 1.** Schematic illustration for synthesis, etching and transfer of large-area graphene films. Transferring and patterning of graphene films grown on a metal/SiO$_2$/Si wafer. Graphene/metal layers supported by polymer films are mechanically separated from a SiO$_2$/Si wafer. After fast etching of metal, the graphene films can be transferred to arbitrary substrates and then patterned using conventional lithography.

**Figure 2.** Photographs of as-grown and transferred wafer scale graphene films. (a) A image of as-grown graphene film on 3-inch 300nm-thick Ni on a SiO$_2$/Si substrate. (b) A transferred wafer-scale graphene film on a PET substrate. (c) and (d) The graphene films printed on a flexible PET and a stretchable rubber substrate. (e) The three-element rosette strain gauge pattern on rubber by prepatterning method.

**Figure 3.** Optical characterization of graphene films grown on Cu and Ni layers. (a) Raman spectra of the large-area graphene layers grown on Cu (blue and red lines) and Ni (black line) films, and transferred on PDMS substrate (green line). The excitation wavelength is 514 nm. (b) Optical transmittance of thin and thick graphene films grown on Cu (red line) and Ni (blue line) layers on SiO$_2$/Si substrates.

**Figure 4.** Electrical properties of graphene FETs and strain sensors. (a) Images of Graphene FET arrays (~16,200 devices) fabricated on a 3-inch SiO$_2$/Si wafer. Source and drain electrodes are formed by 100 nm thick Au. Inset denotes an image of the representative transistor. (b) A transfer curve of the transistor whose channel length and width are both 5 μm. (c) Optical image of a precision mechanical stage used to stretch and contract a PDMS sheet. (d) Resistance modulation of the graphene strain sensor.

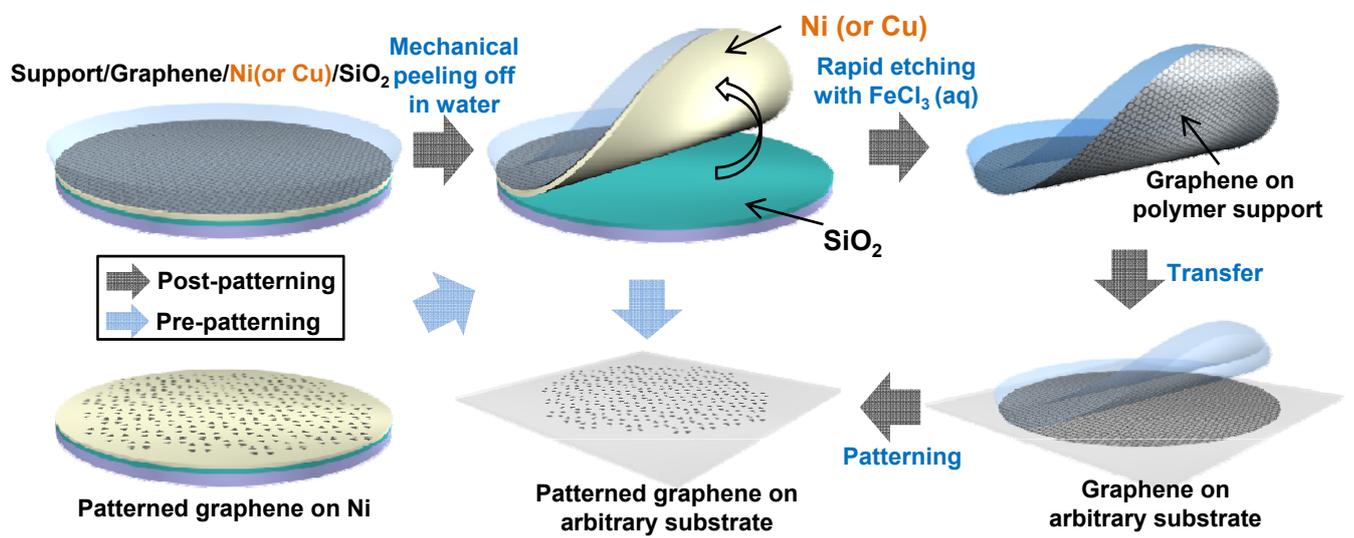

**Fig. 1**

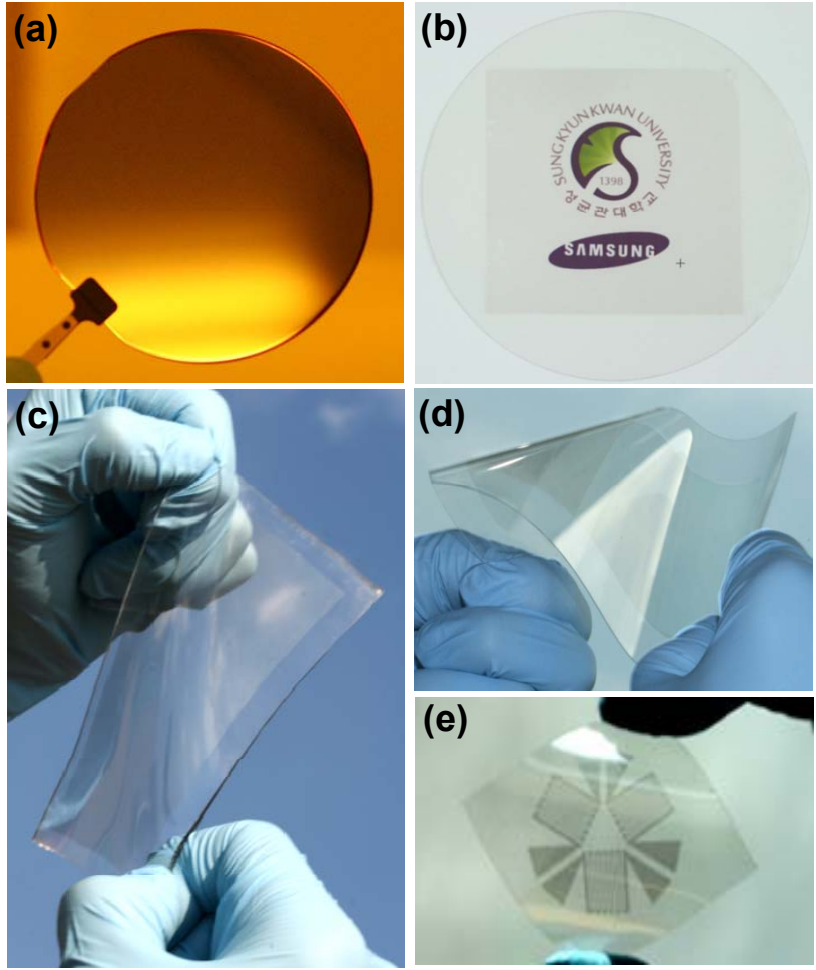

**Fig. 2**

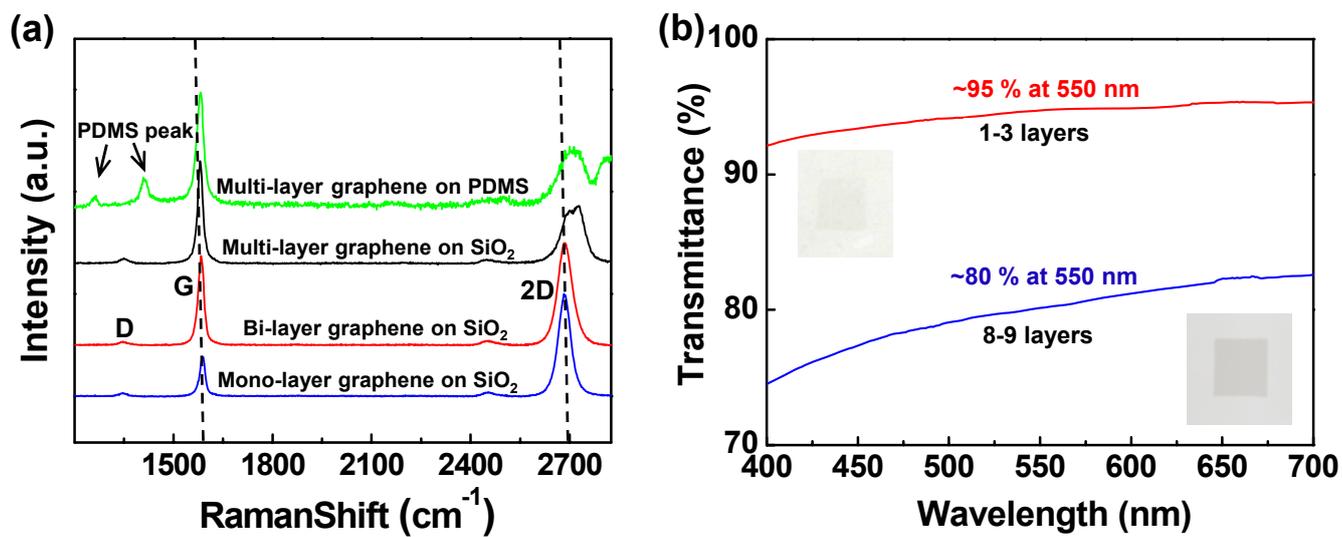

Fig. 3

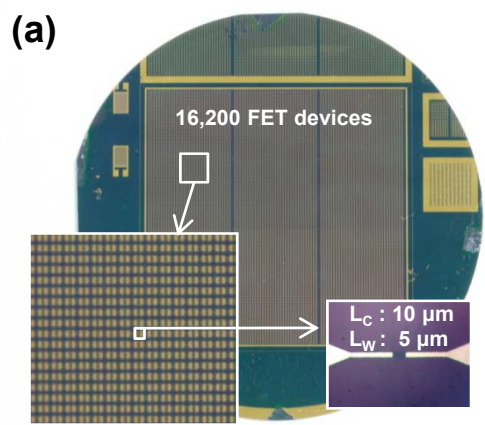
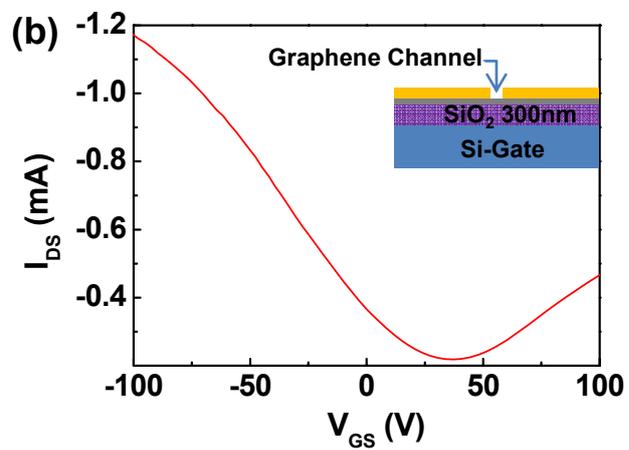
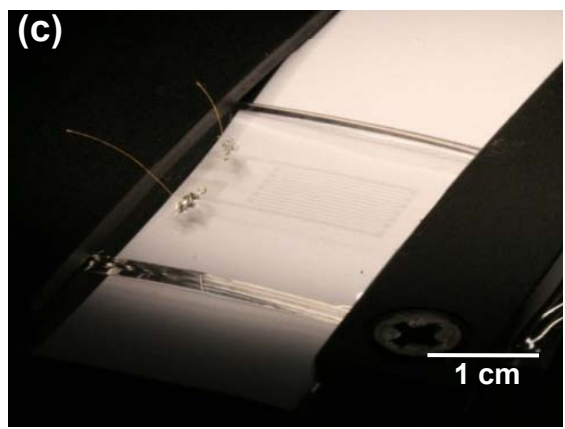
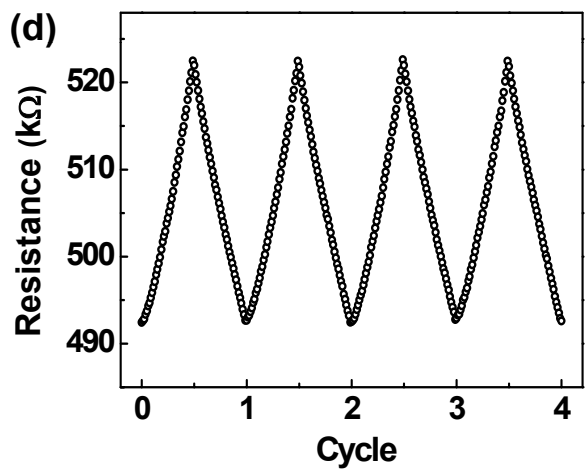

**Fig. 4**